\documentstyle[12pt]{article}
\if@twoside
\else
\fi
\advance\textheight by \topskip
\leftmargin\leftmargini
\labelwidth\leftmargini\advance\labelwidth-\labelsep

\catcode`\@=11
\def\Let@{\relax\iffalse{\fi\let\\=\cr\iffalse}\fi}
\def\vspace@{\def\vspace##1{\crcr\noalign{\vskip##1\relax}}}
\def\multilimits@{\bgroup\vspace@\Let@
 \baselineskip\fontdimen10 \scriptfont\tw@
 \advance\baselineskip\fontdimen12 \scriptfont\tw@
 \lineskip\thr@@\fontdimen8 \scriptfont\thr@@
 \lineskiplimit\lineskip
 \vbox\bgroup\ialign\bgroup\hfil$\m@th\scriptstyle{##}$\hfil\crcr}
\def\Sb{_\multilimits@}
\def\endSb{\crcr\egroup\egroup\egroup}
\def\Sp{^\multilimits@}

\long
\def\QQQ#1#2{}
\def\QTP#1{}
\long
\def\QQA#1#2{}

\def\EXPAND#1[#2]#3{}
\def\NOEXPAND#1[#2]#3{}

\def\LaTeXparent#1{}

\QQQ{Language}{
American English
}

\begin{document}

$ $

\vskip 1cm

\begin{center}
\LARGE{Hamilton-Jacobi approach to Berezinian singular systems} 
\end{center}

\vskip 1cm

\begin{center}
{\large{B. M. Pimentel and R. G. Teixeira}}\\

\vskip 0.5cm

Instituto de F\'{\i}sica Te\'{o}rica\\Universidade Estadual Paulista\\Rua
Pamplona 145\\01405-900 - S\~{a}o Paulo, S.P.\\Brazil\\\thinspace

\vskip 0.5cm

and

\vskip 0.5cm

{\large{J. L. Tomazelli}}\\

\vskip 0.5cm

Departamento de F\'{\i}sica e Qu\'{\i}mica - Faculdade de Engenharia\\%
Universidade Estadual Paulista - Campus de Guaratinguet\'{a}\\Av. Dr.
Ariberto Pereira da Cunha, 333\\12500-000 - Guaratinquet\'{a}, S.P.\\Brazil\\%
\thinspace
\end{center}

\newpage 

$ $

\vskip 6cm

\begin{center}
\begin{minipage}{14.5cm}
\centerline{\bf Abstract}
\,
In this work we present a formal generalization of the 
Hamilton-Jacobi formalism, recently developed for singular systems, to include the 
case of Lagrangians containing variables which are elements of Berezin algebra. 
We derive the Hamilton-Jacobi equation for such systems, analizing the singular 
case in order to obtain the equations of motion as total differential equations 
and study the integrability conditions for such equations. An example is solved 
using both Hamilton-Jacobi and Dirac's Hamiltonian formalisms and the results are 
compared.
\end{minipage}
\end{center}

\newpage\ 

\section{Introduction}

In this work we intend to study singular systems with Lagrangians containing
elements of Berezin algebra from the point of view of the Hamilton-Jacobi
formalism recently developed \cite{Guler 1, Guler 2}. The study of such
systems through Dirac's generalized Hamiltonian formalism has already been
extensively developed in literature \cite{Gitman, Sundermeyer, Regge} and
will be used for comparative purposes.

Despite the success of Dirac's approach in studying singular systems, which
is demonstrated by the wide number of physical systems to which this
formalism has been applied, it is always instructive to study singular
systems through other formalisms, since different procedures will provide
different views for the same problems, even for nonsingular systems. The
Hamilton-Jacobi formalism that we study in this work has been already
generalized to higher order singular systems \cite{Eu 1, Eu 2} and applied
only to a few number of physical examples as the electromagnetic field \cite
{Guler 4}, relativistic particle in an external Electromagnetic field \cite
{Guler 6} and Podolsky's Electrodynamics \cite{Eu 1}. But a better
understanding of this approach utility in the studying singular systems is
still lacking, and such understanding can only be achieved through its
application to other interesting physical systems.

Besides that, Berezin algebra is a useful way to deal simultaneously with
bosonic and fermionic variables in a unique and compact notation, what
justifies the interest in studying systems composed by its elements using
new formalisms.

The aim of this work is not only to generalize the Hamilton-Jacobi approach
for singular systems to the case of Lagrangians containing Berezinian
variables but to present an example of its application to a well known
physical system, comparing the results to those obtained through Dirac's
method.

We will start in section 2 with some basic definitions and next, in section
3, we will introduce the Hamilton-Jacobi formalism to Berezinian systems
using Ca\-ra\-th\'{e}o\-do\-ry's equivalent Lagrangians method. In section 4
the singular case is considered and the equations of motion are obtained as
a system of total differential equations whose integrability conditions are
analyzed in section 5. The equivalence among these integrability conditions
and Dirac's consistency conditions will be discussed separately in the
appendix. In section 6 we present, as an example, the electromagnetic field
coupled to a spinor, which is studied using both the formalism presented in
this work and Dirac's Hamiltonian one. Finally, the conclusions are
presented in section 7.

\section{Basic definitions}

We will start from a Lagrangian $L\left( q,\stackrel{.}{q}\right) $ that
must be an even function of a set of $N$ variables $q^i$ that are elements
of Berezin algebra. For a basic introduction in such algebra we suggest the
reader to refer to ref. \cite{Gitman}, appendix D, from which we took the
definitions used in this paper. A more complete treatment can be found in
ref. \cite{Berezin}.

The Lagrangian equations of motion can be obtained through variational
principles from the action $S=\int $ $Ldt$ 
\begin{equation}
\frac{\delta _rS}{\delta q^i}=\frac{\partial _rL}{\partial q^i}-\frac d{dt}%
\frac{\partial _rL}{\partial \stackrel{.}{q}^i}=0,  \label{a1}
\end{equation}
were we must call attention to the use of right derivatives.

The passage to Hamiltonian formalism is made, as usual, by defining the
momenta variables through {\bf right} derivatives as 
\begin{equation}
p_i\equiv \frac{\partial _rL}{\partial \stackrel{.}{q}^i}  \label{a2}
\end{equation}
and introducing the Hamiltonian function as (summing over repeated indexes) 
\begin{equation}
H=p_i\stackrel{.}{q}^i-L,  \label{a3}
\end{equation}
were the ordering of momenta to the left of velocities shall be observed
since they were defined as right derivatives. This ordering will be, of
course, irrelevant when we deal with even elements of the Berezin algebra.
The Hamiltonian equations of motion will be given by 
\begin{equation}
\stackrel{.}{q}^i=\frac{\partial _lH}{\partial p_i}=\left( -1\right)
^{P_{\left( i\right) }}\frac{\partial _rH}{\partial p_i};\ \stackrel{.}{p}%
_i=-\frac{\partial _rH}{\partial q^i}=-\left( -1\right) ^{P_{\left( i\right)
}}\frac{\partial _lH}{\partial q^i}.  \label{a4}
\end{equation}

If we use the {\it Poisson bracket in the Berezin algebra}, given by 
\begin{equation}
\left\{ F,G\right\} _B=\frac{\partial _rF}{\partial q^i}\frac{\partial _lG}{%
\partial p_i}-\left( -1\right) ^{P_{\left( F\right) }P_{\left( G\right) }}%
\frac{\partial _rG}{\partial q^i}\frac{\partial _lF}{\partial p_i},
\label{a5}
\end{equation}
we get the known expressions 
\begin{equation}
\stackrel{.}{q}^i=\left\{ q^i,H\right\} _B;\ \stackrel{.}{p}_i=\left\{
p_i,H\right\} _B.  \label{a6}
\end{equation}

For simplicity and clarity we will refer to these brackets as {\it Berezin
brackets}. These brackets have similar properties to the usual Poisson
brackets 
\begin{equation}
\left\{ F,G\right\} _B=-\left( -1\right) ^{P_{\left( F\right) }P_{\left(
G\right) }}\left\{ G,F\right\} _B,  \label{a7}
\end{equation}
\begin{equation}
\left\{ F,GK\right\} _B=\left\{ F,G\right\} _BK+\left( -1\right) ^{P_{\left(
F\right) }P_{\left( G\right) }}G\left\{ F,K\right\} _B,  \label{a8}
\end{equation}
\begin{eqnarray}
\left( -1\right) ^{P_{\left( F\right) }P_{\left( K\right) }}\left\{
F,\left\{ G,K\right\} _B\right\} _B&+&\left( -1\right) ^{P_{\left( G\right)
}P_{\left( F\right) }}\left\{ G,\left\{ K,F\right\} _B\right\} _B  \nonumber
\\
&+&\left( -1\right) ^{P_{\left( K\right) }P_{\left( G\right) }}\left\{
K,\left\{ F,G\right\} _B\right\} _B=0,  \label{a9}
\end{eqnarray}
were the last expression is the analogue of Jacobi's identity.

Similarly to the usual case, the transition to phase space is only possible
if the momenta variables, given by Eq. (\ref{a2}), are independent variables
among themselves so that we can express all velocities $\stackrel{.}{q}_i$
as functions of canonical variables $\left( q^i,p_i\right) $. Such necessity
implies that the {\it Hessian supermatrix} 
\begin{equation}
H_{ij}\equiv \frac{\partial _rp_i}{\partial \stackrel{.}{q}^j}=\frac{%
\partial _r^2L}{\partial \stackrel{.}{q}^i\partial \stackrel{.}{q}^j}
\label{a10}
\end{equation}
must be nonsingular. Otherwise, if the Hessian has a rank $P=N-R$, there
will be $R$ relations among the momenta variables and coordinates $q^i$ that
are primary constraints (that we suppose to have definite parity), while $R$
velocities $\stackrel{.}{q}^i$ will remain arbitrary variables in the
theory. The development of Dirac's Generalized Hamiltonian Formalism is
straightforward: the primary constraints have to be added to the
Hamiltonian, we have to work out the consistency conditions, separate the
constraints in first and second-class ones and define the Dirac brackets
using the {\it supermatrix} whose elements are the Poisson brackets among
the second-class constraints \cite{Gitman}.

\section{Hamilton-Jacobi formalism}

From Carath\'{e}odory's equivalent Lagrangians method \cite{Caratheodory} we
can obtain the Ha\-mil\-ton-Jacobi equation for the even Lagrangian $L\left(
q,\stackrel{.}{q}\right) $. The procedure is similar to the one applied to
usual variables: given a Lagrangian $L$, we can obtain a completely
equivalent one given by 
\[
L^{\prime }=L\left( q^i,\stackrel{.}{q}^i\right) -\frac{dS\left(
q^i,t\right) }{dt}, 
\]
were $S\left( q^i,t\right) $ is an even function in order to keep the
equivalent Lagrangian even.

These Lagrangians are equivalent because their respective action integrals
have simultaneous extremum. Then we choose the function $S\left(
q^i,t\right) $ in such a way that we get an extremum of $L^{\prime }$ and,
consequently, an extremum of the Lagrangian $L$.

For this purpose, it is enough to find a set of functions $\beta ^i(q^j,t)$
and $S\left( q^i,t\right) $ such that 
\begin{equation}
L^{\prime }\left( q^i,\stackrel{.}{q}^i=\beta ^i(q^j,t),t\right) =0
\label{a11}
\end{equation}
and for all neighborhood of $\stackrel{.}{q}^i=\beta ^i(q^j,t)$ 
\begin{equation}
L^{\prime }\left( q^i,\stackrel{.}{q}^i\right) >0.  \label{a12}
\end{equation}

With these conditions satisfied, the Lagrangian $L^{\prime }$ (and
consequently $L$) will have a minimum in $\stackrel{.}{q}^i=\beta ^i(q^j,t)$
so that the solutions of the differential equations given by 
\[
\stackrel{.}{q}^i=\beta ^i(q^j,t), 
\]
will correspond to an extremum of the action integral.

From the definition of $L^{\prime }$ we have 
\[
L^{\prime }=L\left( q^j,\stackrel{.}{q}^j\right) -\frac{\partial S\left(
q^j,t\right) }{\partial t}-\frac{\partial _rS\left( q^j,t\right) }{\partial
q^i}\frac{dq^i}{dt}, 
\]
where again we must call attention to the use of the right derivative.

Using condition (\ref{a11}) we have 
\[
\left. \left[ L\left( q^j,\stackrel{.}{q}^j\right) -\frac{\partial S\left(
q^j,t\right) }{\partial t}-\frac{\partial _rS\left( q^j,t\right) }{\partial
q^i}\stackrel{.}{q}^i\right] \right| _{\stackrel{.}{q}^i=\beta ^i}=0, 
\]
\begin{equation}
\left. \frac{\partial S}{\partial t}\right| _{\stackrel{.}{q}^i=\beta
^i}=\left. \left[ L\left( q^j,\stackrel{.}{q}^j\right) -\frac{\partial
_rS\left( q^j,t\right) }{\partial q^i}\stackrel{.}{q}^i\right] \right| _{%
\stackrel{.}{q}^i=\beta ^i}.  \label{a13}
\end{equation}

In addition, since $L^{\prime }$ has a minimum in $\stackrel{.}{q}^i=\beta
^i $, we must have 
\[
\left. \frac{\partial _rL^{\prime }}{\partial \stackrel{.}{q}^i}\right| _{%
\stackrel{.}{q}^i=\beta ^i}=0\Rightarrow \left. \left[ \frac{\partial _rL}{%
\partial \stackrel{.}{q}^i}-\frac{\partial _r}{\partial \stackrel{.}{q}^i}%
\left( \frac{dS}{dt}\right) \right] \right| _{\stackrel{.}{q}^i=\beta ^i}=0, 
\]
\[
\left. \left[ \frac{\partial _rL}{\partial \stackrel{.}{q}^i}-\frac{\partial
_rS\left( q^j,t\right) }{\partial q^i}\right] \right| _{\stackrel{.}{q}%
^i=\beta ^i}=0, 
\]
or 
\begin{equation}
\left. \frac{\partial _rS\left( q^j,t\right) }{\partial q^i}\right| _{%
\stackrel{.}{q}^i=\beta ^i}=\left. \frac{\partial _rL}{\partial \stackrel{.}{%
q}^i}\right| _{\stackrel{.}{q}^i=\beta ^i}.  \label{a14}
\end{equation}

Now, using the definitions for the conjugated momenta given by Eq. (\ref{a2}%
), we get 
\begin{equation}
p_i=\frac{\partial _rS\left( q^j,t\right) }{\partial q^i}.  \label{a15}
\end{equation}

We can see from this result and from Eq. (\ref{a13}) that, in order to
obtaining an extremum of the action, we must get a function $S\left(
q^j,t\right) $ such that 
\begin{equation}
\frac{\partial S}{\partial t}=-H_0  \label{a16}
\end{equation}
where $H_0$ is the canonical Hamiltonian 
\begin{equation}
H_0=p_i\stackrel{.}{q}^i-L\left( q^j,\stackrel{.}{q}^j\right)  \label{a17}
\end{equation}
and the momenta $p_i$ are given by Eq. (\ref{a15}). Besides, Eq. (\ref{a16})
is the Hamilton-Jacobi partial differential equation (HJPDE).

\section{The singular case}

We now consider the case of a system with a singular Lagrangian. When the
Hessian supermatrix is singular with a rank $P=N-R$ we can define the
variables $q^i$ in such order that the $P\times P$ supermatrix in the right
bottom corner of the Hessian supermatrix be nonsingular, i.e. 
\begin{equation}
\det \left| H_{ab}\right| \neq 0;\ H_{ab}=\frac{\partial _r^2L}{\partial 
\stackrel{.}{q}^a\partial \stackrel{.}{q}^b};\ a,b=R+1,...,N.  \label{a18}
\end{equation}

This allows us to solve the velocities $\stackrel{.}{q}^a$ as functions of
coordinates $q^{\prime }s$ and momenta $p_a$, i.e., $\stackrel{.}{q}%
^a=f^a\left( q^i,p_b\right) $.

There will remain $R$ momenta variables $p_\alpha $ dependent upon the other
canonical variables, and we can always \cite{Gitman, Sundermeyer, Mukunda}
write expressions like 
\begin{equation}
p_\alpha =-H_\alpha \left( q^i;\ p_a\right) ;\ \alpha =1,...,R;  \label{a19}
\end{equation}
that correspond to the Dirac's primary constraints $\Phi _\alpha \equiv
p_\alpha +H_\alpha \left( q^i;\ p_a\right) \approx 0$.

The Hamiltonian $H_0$, given by Eq. (\ref{a17}), becomes 
\begin{equation}
H_0=p_af^a+\left. p_\alpha \right| _{p_\beta =-H_\beta }{}\cdot \stackrel{.}{%
q}^\alpha -L\left( q^i,\stackrel{.}{q}_\alpha ,\stackrel{.}{q}_a=f^a\right) ,
\label{a20}
\end{equation}
where $\alpha ,\beta =1,...,R$; $a=R+1,...,N$. On the other hand we have 
\[
\frac{\partial _rH_0}{\partial \stackrel{.}{q}^\alpha }=p_a\frac{\partial
_rf^a}{\partial \stackrel{.}{q}^\alpha }+p_\alpha -\frac{\partial _rL}{%
\partial \stackrel{.}{q}^\alpha }-\frac{\partial _rL}{\partial \stackrel{.}{q%
}^a}\frac{\partial _rf^a}{\partial \stackrel{.}{q}^\alpha }=0, 
\]
so the Hamiltonian $H_0$ does not depend explicitly upon the velocities $%
\stackrel{.}{q}_\alpha $.

Now we will adopt the following notation: the time parameter $t$ will be
called $t^0\equiv q^0$; the coordinates $q^\alpha $ will be called $t^\alpha 
$; the momenta $p_\alpha $ will be called $P_\alpha $ and the {\it momentum} 
$p_0\equiv P_0$ will be {\bf defined} as 
\begin{equation}
P_0\equiv \frac{\partial S}{\partial t}.  \label{a21}
\end{equation}

Then, to get an extremum of the action integral, we must find a function $%
S\left( t_\alpha ;q^a,t\right) $ that satisfies the following set of HJPDE 
\begin{equation}
H_0^{\prime }\equiv P_0+H_0\left( t,t^\beta ;q^a;p_a=\frac{\partial _rS}{%
\partial q^a}\right) =0,  \label{a22}
\end{equation}
\begin{equation}
H_\alpha ^{\prime }\equiv P_\alpha +H_\alpha \left( t,t^\beta ;q^a;p_a=\frac{%
\partial _rS}{\partial q^a}\right) =0.  \label{a23}
\end{equation}
where $\alpha ,\beta =1,...,R$. If we let the indexes $\alpha $ and $\beta $
run from $0$ to $R$ we can write both equations as 
\begin{equation}
H_\alpha ^{\prime }\equiv P_\alpha +H_\alpha \left( t^\beta ;q^a;p_a=\frac{%
\partial _rS}{\partial q^a}\right) =0.  \label{a24}
\end{equation}

From the above definition and Eq. (\ref{a20}) we have 
\[
\frac{\partial _rH_0^{\prime }}{\partial p_b}=-\frac{\partial _rL}{\partial 
\stackrel{.}{q}^a}\frac{\partial _rf^a}{\partial p_b}-\left( -1\right)
^{P_{\left( b\right) }P_{\left( \alpha \right) }}\frac{\partial _rH_\alpha }{%
\partial p_b}\stackrel{.}{q}^\alpha +p_a\frac{\partial _rf^a}{\partial p_b}%
+\left( -1\right) ^{P_{\left( b\right) }}\stackrel{.}{q}^b, 
\]
\[
\frac{\partial _rH_0^{\prime }}{\partial p_b}=\left( -1\right) ^{P_{\left(
b\right) }}\stackrel{.}{q}^b-\left( -1\right) ^{P_{\left( b\right)
}P_{\left( \alpha \right) }}\frac{\partial _rH_\alpha }{\partial p_b}%
\stackrel{.}{q}^\alpha , 
\]
where we came back to $\alpha =1,...,R$.

Multiplying this equation by $dt=dt^0$ and $\left( -1\right) ^{P_{\left(
b\right) }}$, we have 
\[
dq^b=\left( -1\right) ^{P_{\left( b\right) }}\frac{\partial _rH_0^{\prime }}{%
\partial p_b}dt^0+\left( -1\right) ^{P_{\left( b\right) }+P_{\left( b\right)
}P_{\left( \alpha \right) }}\frac{\partial _rH_\alpha ^{\prime }}{\partial
p_b}dq^\alpha . 
\]

Using $t^\alpha \equiv q^\alpha $, letting the index $\alpha $ run again
from $0$ to $R$ and considering $P_{\left( 0\right) }=P_{\left( t^0\right)
}=0$ we have 
\begin{equation}
dq^b=\left( -1\right) ^{P_{\left( b\right) }+P_{\left( b\right) }P_{\left(
\alpha \right) }}\frac{\partial _rH_\alpha ^{\prime }}{\partial p_b}%
dt^\alpha .  \label{a25}
\end{equation}

Noticing that we have the expressions 
\[
dq^\beta =\left( -1\right) ^{P_{\left( \beta \right) }+P_{\left( \beta
\right) }P_{\left( \alpha \right) }}\frac{\partial _rH_\alpha ^{\prime }}{%
\partial p_\beta }dt^\alpha =\left( -1\right) ^{P_{\left( \beta \right)
}+P_{\left( \beta \right) }P_{\left( \alpha \right) }}\delta _\alpha ^\beta
dt^\alpha \equiv dt^\beta 
\]
identically satisfied for $\alpha ,\ \beta =0,1,...,R$, we can write Eq. (%
\ref{a25}) as 
\begin{equation}
dq^i=\left( -1\right) ^{P_{\left( i\right) }+P_{\left( i\right) }P_{\left(
\alpha \right) }}\frac{\partial _rH_\alpha ^{\prime }}{\partial p_i}%
dt^\alpha ;\ i=0,1,...,N.  \label{a26}
\end{equation}

If we consider that we have a solution $S\left( q^j,t\right) $ of the HJPDE
given by Eq. (\ref{a24}) then, differentiating that equation with respect to 
$q^i$, we obtain 
\begin{equation}
\frac{\partial _rH_\alpha ^{\prime }}{\partial q^i}+\frac{\partial
_rH_\alpha ^{\prime }}{\partial P_\beta }\frac{\partial _r^2S}{\partial
t^\beta \partial q^i}+\frac{\partial _rH_\alpha ^{\prime }}{\partial p_a}%
\frac{\partial _r^2S}{\partial q^a\partial q^i}=0  \label{a27}
\end{equation}
for $\alpha ,\ \beta =0,1,...,R$.

From the momenta definitions we can obtain 
\begin{equation}
dp_i=\frac{\partial _r^2S}{\partial q^i\partial t^\beta }dt^\beta +\frac{%
\partial _r^2S}{\partial q^i\partial q^a}dq^a;\ i=0,1,...,N.  \label{a28}
\end{equation}

Now, contracting equation (\ref{a27}) with $dt^\alpha $ (from the right),
multiplying by $\left( -1\right) ^{P_{\left( i\right) }P_{\left( \alpha
\right) }}$ and adding the result to equation (\ref{a28}) we get 
\begin{eqnarray*}
&&dp_i+\left( -1\right) ^{P_{\left( i\right) }P_{\left( \alpha \right) }}%
\frac{\partial _rH_\alpha ^{\prime }}{\partial q^i}dt^\alpha = \\
&&=\left[ \frac{\partial _r^2S}{\partial q^i\partial q^a}\left( dq^a-\left(
-1\right) ^{P_{\left( i\right) }P_{\left( \alpha \right) }}\left( -1\right)
^{\left( P_{\left( \alpha \right) }+P_{\left( a\right) }\right) \left(
P_{\left( i\right) }+P_{\left( a\right) }\right) }\left( -1\right)
^{P_{\left( i\right) }P_{\left( a\right) }}\frac{\partial _rH_\alpha
^{\prime }}{\partial p_a}dt^\alpha \right) \right. \\
&&\left. +\frac{\partial _r^2S}{\partial q^i\partial t^\beta }\left(
dt^\beta -\left( -1\right) ^{P_{\left( i\right) }P_{\left( \alpha \right)
}}\left( -1\right) ^{\left( P_{\left( \alpha \right) }+P_{\left( \beta
\right) }\right) \left( P_{\left( i\right) }+P_{\left( \beta \right)
}\right) }\left( -1\right) ^{P_{\left( i\right) }P_{\left( \beta \right) }}%
\frac{\partial _rH_\alpha ^{\prime }}{\partial P_\beta }dt^\alpha \right)
\right] ,
\end{eqnarray*}
\begin{eqnarray}
dp_i &+&\left( -1\right) ^{P_{\left( i\right) }P_{\left( \alpha \right) }}%
\frac{\partial _rH_\alpha ^{\prime }}{\partial q^i}dt^\alpha =\left[ \frac{%
\partial _r^2S}{\partial q^i\partial q^a}\left( dq^a-\left( -1\right)
^{P_{\left( \alpha \right) }P_{\left( a\right) }+P_{\left( a\right) }}\frac{%
\partial _rH_\alpha ^{\prime }}{\partial p_a}dt^\alpha \right) +\right. 
\nonumber \\
&&\left. +\frac{\partial _r^2S}{\partial q^i\partial t^\beta }\left(
dt^\beta -\left( -1\right) ^{P_{\left( \alpha \right) }P_{\left( \beta
\right) }+P_{\left( \beta \right) }}\frac{\partial _rH_\alpha ^{\prime }}{%
\partial P_\beta }dt^\alpha \right) \right] ,  \label{a29}
\end{eqnarray}
where we used the fact that 
\[
\frac{\partial _r^2S}{\partial q^j\partial q^i}=\left( -1\right) ^{P_{\left(
i\right) }P_{\left( j\right) }}\frac{\partial _r^2S}{\partial q^i\partial q^j%
}, 
\]
and that we have the following parities 
\[
P\left( \frac{\partial _r^2S}{\partial q^i\partial q^j}\right) =P_{\left(
i\right) }+P_{\left( j\right) }, 
\]
\[
P\left( \frac{\partial _rH_\alpha ^{\prime }}{\partial p_j}\right)
=P_{\left( \alpha \right) }+P_{\left( j\right) }. 
\]

If the total differential equation given by Eq. (\ref{a26}) applies, the
above equation becomes 
\begin{equation}
dp_i=-\left( -1\right) ^{P_{\left( i\right) }P_{\left( \alpha \right) }}%
\frac{\partial _rH_\alpha ^{\prime }}{\partial q^i}dt^\alpha ;\ i=0,1,...,N.
\label{a30}
\end{equation}

Making $Z\equiv S\left( t^\alpha ;q^a\right) $ and using the momenta
definitions together with Eq. (\ref{a26}) we have 
\[
dZ=\frac{\partial _rS}{\partial t^\beta }dt^\beta +\frac{\partial _rS}{%
\partial q^a}dq^a, 
\]

\[
dZ=-H_\beta dt^\beta +p_a\left( \left( -1\right) ^{P_{\left( a\right)
}+P_{\left( a\right) }P_{\left( \alpha \right) }}\frac{\partial _rH_\alpha
^{\prime }}{\partial p_a}dt^\alpha \right) . 
\]

With a little change of indexes we get 
\begin{eqnarray}
dZ=\left( -H_\beta +\left( -1\right) ^{P_{\left( a\right) }+P_{\left(
a\right) }P_{\left( \beta \right) }}p_a\frac{\partial _rH_\beta ^{\prime }}{%
\partial p_a}\right) dt^\beta .  \label{a31}
\end{eqnarray}

This equation together with Eq. (\ref{a26}) and Eq. (\ref{a30}) are the
total differential equations for the characteristics curves of the HJPDE
given by Eq. (\ref{a24}) and, if they form a completely integrable set,
their simultaneous solutions determine $S\left( t^\alpha ;q^a\right) $
uniquely from the initial conditions.

Besides that, Eq. (\ref{a26}) and Eq. (\ref{a30}) are the equations of
motion of the system written as total differential equations. It is
important to observe that, in the nonsingular case, we have only $%
H_0^{\prime }\neq 0$ and no others $H_\alpha ^{\prime }$; so that these
equations of motion will reduce naturally to the usual expressions given by
Eq. (\ref{a4}).

\section{Integrability conditions}

The analysis of integrability conditions of the total differential equations
(\ref{a26}), (\ref{a30}) and (\ref{a31}) can be carried out using standard
techniques. This have already been made \cite{Guler 2, Guler 3, Guler 5} for
systems with usual variables, and here we will present the analysis of the
integrability conditions for Berezinian singular systems.

To a given set of total differential equations 
\begin{equation}
dq^i=\Lambda ^i{}_\alpha \left( t^\beta ,q^j\right) dt^\alpha  \label{i1}
\end{equation}
($i,\ j=0,1,...,N$ and $\alpha ,\ \beta =0,1,...,R<N$) we may associate a
set of partial differential equations \cite{Whittaker} 
\begin{equation}
X_\alpha f=\frac{\partial _rf}{\partial q^i}\Lambda ^i{}_\alpha =0,
\label{i2}
\end{equation}
where $X_\alpha $ are linear operators.

Given any twice differentiable solution of the set (\ref{i2}), it should
also satisfy the equation 
\begin{equation}
\left[ X_\alpha ,X_\beta \right] f=0,  \label{i4}
\end{equation}
where 
\begin{equation}
\left[ X_\alpha ,X_\beta \right] f=\left( X_\alpha X_\beta -\left( -1\right)
^{P_{\left( \alpha \right) }P_{\left( \beta \right) }}X_\beta X_\alpha
\right) f  \label{i5}
\end{equation}
is the bracket among the operators $X_\alpha $. This implies that we should
have 
\begin{equation}
\left[ X_\alpha ,X_\beta \right] f=C_{\alpha \beta }{}^\gamma X_\gamma f.
\label{i6}
\end{equation}

So, the commutation relations (\ref{i5}) will give the maximal number of
linearly independent equations. Any commutator that results in a expression
that can't be written as Eq. (\ref{i6}) must be written as a new operator $X$
and be joined to the original set (\ref{i2}), having all commutators with
the other $X^{\prime }s$ calculated. The process is repeated until all
operators $X$ satisfy Eq. (\ref{i6}).

If all operators $X_\alpha $ satisfy the commutation relations given by Eq. (%
\ref{i6}) the system of partial differential equations (\ref{i2}) is said to
be {\it complete} and the corresponding system of total differential
equations (\ref{i1}) is integrable if, and only if, the system (\ref{i2}) is
complete.

Now, we consider the system of differential equations obtained in the
previous section. First we shall observe that if the total differential
equations (\ref{a26}) and (\ref{a30}) are integrable the solutions of Eq. (%
\ref{a31}) can be obtained by a quadrature, so we only need to analyze the
integrability conditions for the last ones, since the former will be
integrable as a consequence.

The operators $X_\alpha $ corresponding to the system of total differential
equations formed by Eq. (\ref{a26}) and Eq. (\ref{a30}) are given by, 
\[
X_\alpha f\left( t^\beta ,q^a,p^a\right) =\frac{\partial _rf}{\partial q^i}%
\left( -1\right) ^{P_{\left( i\right) }+P_{\left( i\right) }P_{\left( \alpha
\right) }}\frac{\partial _rH_\alpha ^{\prime }}{\partial p_i}-\frac{\partial
_rf}{\partial p_i}\left( -1\right) ^{P_{\left( i\right) }P_{\left( \alpha
\right) }}\frac{\partial _rH_\alpha ^{\prime }}{\partial q^i}, 
\]
\begin{eqnarray*}
&&\hspace{-1cm}X_\alpha f\left( t^\beta ,q^a,p^a\right) =\frac{\partial _rf}{%
\partial q^i}\frac{\partial _lH_\alpha ^{\prime }}{\partial p_i} \\
&&-\left( -1\right) ^{P_{\left( i\right) }+P_{\left( i\right) }P_{\left(
f\right) }}\left( -1\right) ^{P_{\left( i\right) }P_{\left( \alpha \right)
}}\left( -1\right) ^{\left( P_{\left( \alpha \right) }+P_{\left( i\right)
}\right) \left( P_{\left( f\right) }+P_{\left( i\right) }\right) }\frac{%
\partial _rH_\alpha ^{\prime }}{\partial q^i}\frac{\partial _lf}{\partial p_i%
},
\end{eqnarray*}
\[
X_\alpha f\left( t^\beta ,q^a,p^a\right) =\frac{\partial _rf}{\partial q^i}%
\frac{\partial _lH_\alpha ^{\prime }}{\partial p_i}-\left( -1\right)
^{P_{\left( \alpha \right) }P_{\left( f\right) }}\frac{\partial _rH_\alpha
^{\prime }}{\partial q^i}\frac{\partial _lf}{\partial p_i}, 
\]
\begin{equation}
X_\alpha f\left( t^\beta ,q^a,p^a\right) =\left\{ f,H_\alpha ^{\prime
}\right\} _B;  \label{i7}
\end{equation}
where $i=0,1,...,N$; $\alpha ,\ \beta =0,1,...,R$; $a=R+1,...,N$ and we have
used Eq. (\ref{i1}) and Eq. (\ref{i2}) together with the result 
\begin{equation}
\frac{\partial _lA}{\partial q^i}=\left( -1\right) ^{P_{\left( i\right)
}}\left( -1\right) ^{P_{\left( i\right) }P_{\left( A\right) }}\frac{\partial
_rA}{\partial q^i}.  \label{i8}
\end{equation}

It is important to notice that the Berezin bracket in Eq. (\ref{i7}) is
defined in a $2N+2$ dimensional phase space, since we are including $q^0=t$
as a ``coordinate''.

Now, the integrability condition will be 
\begin{eqnarray*}
\left[ X_\alpha ,X_\beta \right] f &=&\left( X_\alpha X_\beta -\left(
-1\right) ^{P_{\left( \alpha \right) }P_{\left( \beta \right) }}X_\beta
X_\alpha \right) f \\
&=&X_\alpha \left\{ f,H_\beta ^{\prime }\right\} _B-\left( -1\right)
^{P_{\left( \alpha \right) }P_{\left( \beta \right) }}X_\beta \left\{
f,H_\alpha ^{\prime }\right\} _B=0,
\end{eqnarray*}
\begin{equation}
\left[ X_\alpha ,X_\beta \right] f=\left\{ \left\{ f,H_\beta ^{\prime
}\right\} _B,H_\alpha ^{\prime }\right\} _B-\left( -1\right) ^{P_{\left(
\alpha \right) }P_{\left( \beta \right) }}\left\{ \left\{ f,H_\alpha
^{\prime }\right\} _B,H_\beta ^{\prime }\right\} _B=0,  \label{i9}
\end{equation}
that will reduce to 
\begin{eqnarray*}
\left[ X_\alpha ,X_\beta \right] f &=&-\left( -1\right) ^{\left( P_{\left(
f\right) }+P_{\left( \beta \right) }\right) P_{\left( \alpha \right)
}}\left\{ H_\alpha ^{\prime },\left\{ f,H_\beta ^{\prime }\right\}
_B\right\} _B \\
&&-\left( -1\right) ^{P_{\left( \alpha \right) }P_{\left( \beta \right)
}}\left( -1\right) ^{\left( P_{\left( f\right) }+P_{\left( \alpha \right)
}\right) P_{\left( \beta \right) }+P_{\left( f\right) }P_{\left( \alpha
\right) }}\left\{ H_\beta ^{\prime },\left\{ H_\alpha ^{\prime },f\right\}
_B\right\} _B=0,
\end{eqnarray*}
\begin{eqnarray*}
\left[ X_\alpha ,X_\beta \right] f &=&-\left( -1\right) ^{P_{\left( f\right)
}P_{\left( \alpha \right) }}\left( -1\right) ^{P_{\left( \beta \right)
}P_{\left( \alpha \right) }}\left\{ H_\alpha ^{\prime },\left\{ f,H_\beta
^{\prime }\right\} _B\right\} _B \\
&&-\left( -1\right) ^{P_{\left( \alpha \right) }P_{\left( \beta \right)
}}\left( -1\right) ^{P_{\left( f\right) }P_{\left( \beta \right) }+P_{\left(
\alpha \right) }P_{\left( \beta \right) }+P_{\left( f\right) }P_{\left(
\alpha \right) }}\left\{ H_\beta ^{\prime },\left\{ H_\alpha ^{\prime
},f\right\} _B\right\} _B=0,
\end{eqnarray*}
\begin{eqnarray*}
\left[ X_\alpha ,X_\beta \right] f &=&-\left( -1\right) ^{P_{\left( f\right)
}P_{\left( \alpha \right) }}\left[ \left( -1\right) ^{P_{\left( \beta
\right) }P_{\left( \alpha \right) }}\left\{ H_\alpha ^{\prime },\left\{
f,H_\beta ^{\prime }\right\} _B\right\} _B\right. \\
&&\left. +\left( -1\right) ^{P_{\left( f\right) }P_{\left( \beta \right)
}}\left\{ H_\beta ^{\prime },\left\{ H_\alpha ^{\prime },f\right\}
_B\right\} _B\right] =0,
\end{eqnarray*}
\[
\left[ X_\alpha ,X_\beta \right] f=-\left( -1\right) ^{P_{\left( f\right)
}P_{\left( \alpha \right) }}\left[ -\left( -1\right) ^{P_{\left( f\right)
}P_{\left( \alpha \right) }}\left\{ f,\left\{ H_\beta ^{\prime },H_\alpha
^{\prime }\right\} _B\right\} _B\right] , 
\]
\begin{equation}
\left[ X_\alpha ,X_\beta \right] f=\left\{ f,\left\{ H_\beta ^{\prime
},H_\alpha ^{\prime }\right\} _B\right\} _B=0,  \label{i10}
\end{equation}
when using the Jacobi relations for Berezin brackets given by Eq. (\ref{a9})
and the fact that 
\begin{equation}
P\left( \left\{ A,B\right\} _B\right) =P\left( A\right) +P\left( B\right) .
\label{i11}
\end{equation}

So, the integrability condition will be 
\begin{equation}
\left\{ H_\beta ^{\prime },H_\alpha ^{\prime }\right\} _B=0;\ \forall \alpha
,\beta .  \label{i12}
\end{equation}

It is important to notice that the above condition can be shown to be
equivalent to the consistency conditions in Dirac's Hamiltonian formalism
but, to keep the continuity of the presentation, we will postpone the
demonstration of this equivalence to the appendix.

Now, the total differential for any function $F\left( t^\beta
,q^a,p^a\right) $ can be written as 
\begin{equation}
dF=\frac{\partial _rF}{\partial q^a}dq^a+\frac{\partial _rF}{\partial p^a}%
dp^a+\frac{\partial _rF}{\partial t^\alpha }dt^\alpha ,  \label{i12a}
\end{equation}
\[
dF=\frac{\partial _rF}{\partial q^a}\left( -1\right) ^{P_{\left( a\right)
}+P_{\left( a\right) }P_{\left( \alpha \right) }}\frac{\partial _rH_\alpha
^{\prime }}{\partial p_a}dt^\alpha -\frac{\partial _rF}{\partial p^a}\left(
-1\right) ^{P_{\left( a\right) }P_{\left( \alpha \right) }}\frac{\partial
_rH_\alpha ^{\prime }}{\partial q^a}dt^\alpha +\frac{\partial _rF}{\partial
t^\alpha }dt^\alpha , 
\]
\[
dF=\left( \frac{\partial _rF}{\partial q^a}\frac{\partial _lH_\alpha
^{\prime }}{\partial p_a}-\left( -1\right) ^{P_{\left( a\right) }}\left(
-1\right) ^{P_{\left( a\right) }P_{\left( F\right) }}\frac{\partial _lF}{%
\partial p^a}\left( -1\right) ^{P_{\left( a\right) }P_{\left( \alpha \right)
}}\frac{\partial _rH_\alpha ^{\prime }}{\partial q^a}+\frac{\partial _rF}{%
\partial t^\alpha }\right) dt^\alpha , 
\]
\[
dF=\left( \frac{\partial _rF}{\partial q^a}\frac{\partial _lH_\alpha
^{\prime }}{\partial p_a}-\left( -1\right) ^{P_{\left( \alpha \right)
}P_{\left( F\right) }}\frac{\partial _rH_\alpha ^{\prime }}{\partial q^a}%
\frac{\partial _lF}{\partial p^a}+\frac{\partial _rF}{\partial t^\alpha }%
\right) dt^\alpha , 
\]
\begin{equation}
dF=\left\{ F,H_\alpha ^{\prime }\right\} _Bdt^\alpha ;  \label{i13}
\end{equation}
where the Berezin bracket above is the one defined in the $2N+2$ phase space
used in Eq. (\ref{i7}). Using this result we have 
\begin{equation}
dH_\beta ^{\prime }=\left\{ H_\beta ^{\prime },H_\alpha ^{\prime }\right\}
_Bdt^\alpha  \label{i14}
\end{equation}
and, consequently, the integrability condition (\ref{i12}) reduces to 
\begin{equation}
dH_\alpha ^{\prime }=0,\ \forall \alpha .  \label{i15}
\end{equation}

If the above conditions are not identically satisfied we will have one of
two different cases. First, we may have a new $H^{\prime }=0$, which has to
satisfy a condition $dH^{\prime }=0$, and must be used in all equations.
Otherwise we will have relations among the differentials $dt^\alpha $ which
also must be used in the remaining equations of the formalism.

\section{Example}

As an example we analyze the case of the electromagnetic field coupled to a
spinor, whose Hamiltonian formalism was analyzed in references \cite{Gitman,
Sundermeyer}. We will consider the Lagrangian density written as 
\begin{equation}
{\cal L}=-\frac 14F_{\mu \nu }F^{\mu \nu }+i\stackrel{\_}{\psi }\gamma ^\mu
\left( \partial _\mu +ieA_\mu \right) \psi -m\stackrel{\_}{\psi }\psi ,
\label{a32}
\end{equation}
where $A_\mu $ are even variables while $\psi $ and $\stackrel{\_}{\psi }$
are odd ones. The electromagnetic tensor is defined as $F^{\mu \nu
}=\partial ^\mu A^\nu -\partial ^\nu A^\mu $ and we are adopting the
Minkowski metric $\eta _{\mu \nu }=diag(+1,-1,-1,-1)$.

\subsection{Hamiltonian formalism}

Let's first review Dirac's Hamiltonian formalism. The momenta variables
conjugated, respectively, to $A_\mu $, $\psi $ and $\stackrel{\_}{\psi }$,
are 
\begin{equation}
p_\mu =\frac{\partial _r{\cal L}}{\partial \stackrel{.}{A}^\mu }=-F_{0\mu },
\label{a33}
\end{equation}
\begin{equation}
p_\psi =\frac{\partial _r{\cal L}}{\partial \stackrel{.}{\psi }}=i\stackrel{%
\_}{\psi }\gamma ^0,\ p_{\stackrel{\_}{\psi }}=\frac{\partial _r{\cal L}}{%
\partial \stackrel{.}{\stackrel{\_}{\psi }}}=0,  \label{a34}
\end{equation}
where we must call attention to the necessity of being careful with the
spinor indexes. Considering, as usual, $\psi $ as a column vector and $%
\stackrel{\_}{\psi }$ as a row vector implies that $p_\psi $ will be a row
vector while $p_{\stackrel{\_}{\psi }}$ will be a column vector.

From the momenta expressions we have the primary constraints 
\begin{equation}
\phi _1=p_0\approx 0,  \label{a35}
\end{equation}
\begin{equation}
\phi _2=p_\psi -i\stackrel{\_}{\psi }\gamma ^0,\ \phi _3=p_{\stackrel{\_}{%
\psi }}\approx 0.  \label{a36}
\end{equation}

The canonical Hamiltonian is given by 
\begin{equation}
H_C=\int {\cal H}_Cd^3x=\int \left( p_\mu \stackrel{.}{A}^\mu +\left( p_\psi
\right) _\alpha \left( \stackrel{.}{\psi }\right) _\alpha +\left( \ p_{%
\stackrel{\_}{\psi }}\right) _\alpha \left( \stackrel{.}{\stackrel{\_}{\psi }%
}\right) _\alpha -{\cal L}\right) d^3x,  \label{a37}
\end{equation}
\begin{eqnarray}
H_C &=&\int {\cal H}_Cd^3x=\int \left[ \frac 14F_{ij}F^{ij}-\frac
12p_ip^i-A_0\left( \partial _ip^i-e\stackrel{\_}{\psi }\gamma ^0\psi \right)
\right.  \nonumber \\
&&\qquad \qquad \qquad \left. -i\stackrel{\_}{\psi }\gamma ^j\left( \partial
_j+ieA_j\right) \psi +m\stackrel{\_}{\psi }\psi \right] d^3x.  \label{a38}
\end{eqnarray}

The primary Hamiltonian is 
\begin{equation}
H_P=\int \left( {\cal H}_C+\lambda _1\phi _1+\phi _2\lambda _2+\lambda
_3\phi _3\right) d^3x,  \label{a39}
\end{equation}
where $\lambda _1$ is an even variable and $\lambda _2$, $\lambda _3$ are
odd variables, $\lambda _2$ being a column vector and $\lambda _3$ a row
vector. The fundamental nonvanishing Berezin brackets (here the brackets are
the ones defined by Eq. (\ref{a5}) in the $2N$ phase space) are 
\begin{equation}
\left\{ A_\mu \left( x\right) ,p^\nu \left( y\right) \right\} _B=\delta _\mu
{}^\nu \delta ^3\left( x-y\right) ,  \label{a40}
\end{equation}
\begin{equation}
\left\{ \psi \left( x\right) ,p_\psi \left( y\right) \right\} _B=\delta
^3\left( x-y\right) ,  \label{a41}
\end{equation}
\begin{equation}
\left\{ \stackrel{\_}{\psi }\left( x\right) ,p_{\stackrel{\_}{\psi }}\left(
y\right) \right\} _B=\delta ^3\left( x-y\right) .  \label{a42}
\end{equation}

The consistency conditions are 
\begin{equation}
\left\{ \phi _1,H_P\right\} _B=\partial _ip^i-e\stackrel{\_}{\psi }\gamma
^0\psi \approx 0,  \label{a43}
\end{equation}
\begin{equation}
\left\{ \phi _2,H_P\right\} _B=-i\left( \partial _j-ieA_j\right) \stackrel{\_%
}{\psi }\gamma ^j-e\stackrel{\_}{\psi }\gamma ^0A_0-m\stackrel{\_}{\psi }%
+i\lambda _3\gamma ^0\approx 0,  \label{a44}
\end{equation}
\begin{equation}
\left\{ \phi _3,H_P\right\} _B=-i\left( \partial _j+ieA_j\right) \gamma
^j\psi +e\gamma ^0A_0\psi +m\psi -i\gamma ^0\lambda _2\approx 0.  \label{a45}
\end{equation}

The last two ones will determine $\lambda _2$ and $\lambda _3$ while the
first one will give rise to the secondary constraint 
\begin{equation}
\chi =\partial _ip^i-e\stackrel{\_}{\psi }\gamma ^0\psi \approx 0,
\label{a46}
\end{equation}
for which the consistency condition will be identically satisfied with the
use of the expressions for $\lambda _2$ and $\lambda _3$ given by Eq. (\ref
{a44}) and Eq. (\ref{a45}). Taking the Berezin brackets among the
constraints we have as nonvanishing results 
\begin{equation}
\left\{ \left( \phi _2\right) _\alpha ,\left( \phi _3\right) _\beta \right\}
_B=-i\left\{ \left( \stackrel{\_}{\psi }\gamma ^0\right) _\alpha ,\left( p_{%
\stackrel{\_}{\psi }}\right) _\beta \right\} _B=-i\left( \gamma ^0\right)
_{\beta \alpha },  \label{a47}
\end{equation}
\begin{equation}
\left\{ \left( \phi _2\right) _\alpha ,\chi \right\} _B=-e\left\{ \left(
p_\psi \right) _\alpha ,\stackrel{\_}{\psi }\gamma ^0\psi \right\}
_B=e\left( \stackrel{\_}{\psi }\gamma ^0\right) _\alpha ,  \label{a48}
\end{equation}
\begin{equation}
\left\{ \left( \phi _3\right) _\alpha ,\chi \right\} _B=-e\left\{ \left( p_{%
\stackrel{\_}{\psi }}\right) _\alpha ,\stackrel{\_}{\psi }\gamma ^0\psi
\right\} _B=-e\left( \gamma ^0\psi \right) _\alpha ,  \label{a49}
\end{equation}
where we explicitly wrote the spinor indexes. Obviously the $\phi _1$
constraint is first class, but we have another first class constraint. This
can be seen from the supermatrix $\Delta $ formed by the Berezin brackets
among the second class constraints $\chi $, $\phi _2$ and $\phi _3$.
Numbering the constraints as $\Phi _1=\chi $, $\Phi _2=\phi _2$ and $\Phi
_3=\phi _3$ we have this supermatrix in normal form (see ref.\cite{Gitman},
appendix D) given, with spinor indexes indicated, by 
\begin{equation}
\Delta =\left( 
\begin{array}{ccc}
\left\{ \Phi _1,\Phi _1\right\} _B & \left\{ \Phi _1,\left( \Phi _2\right)
_\alpha \right\} _B & \left\{ \Phi _1,\left( \Phi _3\right) _\beta \right\}
_B \\ 
\left\{ \left( \Phi _2\right) _\delta ,\Phi _1\right\} _B & \left\{ \left(
\Phi _2\right) _\delta ,\left( \Phi _2\right) _\alpha \right\} _B & \left\{
\left( \Phi _2\right) _\delta ,\left( \Phi _3\right) _\beta \right\} _B \\ 
\left\{ \left( \Phi _3\right) _\mu ,\Phi _1\right\} _B & \left\{ \left( \Phi
_3\right) _\mu ,\left( \Phi _2\right) _\alpha \right\} _B & \left\{ \left(
\Phi _3\right) _\mu ,\left( \Phi _3\right) _\beta \right\} _B
\end{array}
\right) ,  \label{a50}
\end{equation}
\begin{equation}
\Delta =\left( 
\begin{array}{ccc}
0 & -e\left( \stackrel{\_}{\psi }\gamma ^0\right) _\alpha & e\left( \gamma
^0\psi \right) _\beta \\ 
e\left( \stackrel{\_}{\psi }\gamma ^0\right) _\delta & 0 & -i\left( \gamma
^0\right) _{\beta \delta } \\ 
-e\left( \gamma ^0\psi \right) _\mu & -i\left( \gamma ^0\right) _{\mu \alpha
} & 0
\end{array}
\right) .  \label{a51}
\end{equation}

This supermatrix has one eingevector with null eingevalue that is 
\begin{equation}
\left( 
\begin{array}{c}
1 \\ 
ie\left( \stackrel{\_}{\psi }\right) _\alpha \\ 
-ie\left( \psi \right) _\beta
\end{array}
\right) ,  \label{a52}
\end{equation}
so there is another first class constraint given by 
\begin{equation}
\varphi =\chi +ie\left( \Phi _2\right) _\alpha \left( \stackrel{\_}{\psi }%
\right) _\alpha -ie\left( \Phi _3\right) _\beta \left( \psi \right) _\beta ,
\label{a53}
\end{equation}
\begin{equation}
\varphi =\partial _ip^i+ie\left( p_\psi \psi +\stackrel{\_}{\psi }p_{%
\stackrel{\_}{\psi }}\right) ,  \label{a54}
\end{equation}
that we will substitute for $\chi $. So, we have the first class constraints 
$\phi _1$ and $\varphi $, and the second class ones $\phi _2$ and $\phi _3$.
The supermatrix $\Delta $ now reduces to the Berezin brackets among the
second class constraints $\phi _2$ and $\phi _3$ and is given by 
\begin{equation}
\Delta =\left( 
\begin{array}{cc}
\left\{ \left( \Phi _2\right) _\delta ,\left( \Phi _2\right) _\alpha
\right\} _B & \left\{ \left( \Phi _2\right) _\delta ,\left( \Phi _3\right)
_\beta \right\} _B \\ 
\left\{ \left( \Phi _3\right) _\mu ,\left( \Phi _2\right) _\alpha \right\} _B
& \left\{ \left( \Phi _3\right) _\mu ,\left( \Phi _3\right) _\beta \right\}
_B
\end{array}
\right) =\left( 
\begin{array}{cc}
0 & -i\left( \gamma ^0\right) _{\beta \delta } \\ 
-i\left( \gamma ^0\right) _{\mu \alpha } & 0
\end{array}
\right) ,  \label{a55}
\end{equation}
having as inverse 
\begin{equation}
\Delta ^{-1}=\left( 
\begin{array}{cc}
0 & i\left( \gamma ^0\right) _{\alpha \lambda } \\ 
i\left( \gamma ^0\right) _{\delta \gamma } & 0
\end{array}
\right) .  \label{a56}
\end{equation}

With these result, the Dirac brackets among any variables $F$ and $G$ are 
\begin{eqnarray}
\left\{ F\left( x\right) ,G\left( y\right) \right\} _D &=&\left\{ F\left(
x\right) ,G\left( y\right) \right\} _B  \nonumber \\
&&-i\int d^3z\left( \left\{ F\left( x\right) ,\left( \Phi _2\right) _\alpha
\left( z\right) \right\} _B\left( \gamma ^0\right) _{\alpha \beta }\left\{
\left( \Phi _3\right) _\beta \left( z\right) ,G\left( y\right) \right\}
_B\right.  \nonumber \\
&&+\left. \left\{ F\left( x\right) ,\left( \Phi _3\right) _\alpha \left(
z\right) \right\} _B\left( \gamma ^0\right) _{\beta \alpha }\left\{ \left(
\Phi _2\right) _\beta \left( z\right) ,G\left( y\right) \right\} \right) _B.
\label{a57}
\end{eqnarray}

The nonvanishing fundamental brackets now will be 
\begin{equation}
\left\{ A_\mu \left( x\right) ,p^\nu \left( y\right) \right\} _D=\delta _\mu
\,^\nu \delta ^3\left( x-y\right) ,  \label{a58}
\end{equation}
\begin{equation}
\left\{ \left( \psi \right) _\lambda \left( x\right) ,\left( \stackrel{\_}{%
\psi }\right) _\alpha \left( y\right) \right\} _D=-i\left( \gamma ^0\right)
_{\lambda \alpha }\delta ^3\left( x-y\right) ,  \label{a59}
\end{equation}
\begin{equation}
\left\{ \left( \psi \right) _\lambda \left( x\right) ,\left( p_\psi \right)
_\alpha \left( y\right) \right\} _D=\delta _{\lambda \alpha }\delta ^3\left(
x-y\right) .  \label{a60}
\end{equation}

Now we can make the second class constraints as strong equalities and write
the equations of motion in terms of the Dirac brackets and the extended
Hamiltonian given by 
\begin{equation}
H_E=\int {\cal H}_Ed^3x=\int \left( {\cal H}_C+\lambda _1\phi _1+\alpha
\varphi \right) d^3x.  \label{a61}
\end{equation}

We must remember that, when making $\phi _2=\phi _3\equiv 0$ the constraint $%
\varphi $ becomes identical to the original secondary constraint $\chi $.
Then, the equations of motion will be 
\begin{equation}
\stackrel{.}{A}^i\approx \left\{ A^i,H_E\right\} _D=-p^i+\partial
^iA_0-\partial ^i\alpha ,  \label{a62}
\end{equation}
\begin{equation}
\stackrel{.}{A}^0\approx \left\{ A^0,H_E\right\} _D=\lambda _1,  \label{a63}
\end{equation}
\begin{equation}
\stackrel{.}{p}^i\approx \left\{ p^i,H_E\right\} _D=\partial _jF^{ji}-e%
\stackrel{\_}{\psi }\gamma ^i\psi ,  \label{a64}
\end{equation}
\begin{equation}
\stackrel{.}{\psi }\approx \left\{ \psi ,H_E\right\} _D=-ieA_0\psi -\gamma
^0\gamma ^j\left( \partial _j+ieA_j\right) \psi -im\gamma ^0\psi ,
\label{a65}
\end{equation}
\begin{equation}
\stackrel{.}{\stackrel{\_}{\psi }}\approx \left\{ \stackrel{\_}{\psi }%
,H_E\right\} _D=ieA_0\stackrel{\_}{\psi }-\left( \partial _j-ieA_j\right) 
\stackrel{\_}{\psi }\gamma ^j\gamma ^0+im\stackrel{\_}{\psi }\gamma ^0.
\label{a66}
\end{equation}

Multiplying Eq. (\ref{a65}) from the left by $i\gamma ^0$ we get 
\begin{equation}
i\gamma ^0\stackrel{.}{\psi }=eA_0\gamma ^0\psi -i\gamma ^j\left( \partial
_j+ieA_j\right) \psi +m\psi ,  \label{a67}
\end{equation}
\begin{equation}
i\left( \partial _\mu +ieA_\mu \right) \gamma ^\mu \psi -m\psi =0,
\label{a68}
\end{equation}
while multiplying Eq. (\ref{a66}) from the right by $i\gamma ^0$ we get 
\begin{equation}
i\stackrel{.}{\stackrel{\_}{\psi }}\gamma ^0=-e\stackrel{\_}{\psi }\gamma
^0A_0-i\left( \partial _j+ieA_j\right) \stackrel{\_}{\psi }\gamma ^j-m%
\stackrel{\_}{\psi },  \label{a69}
\end{equation}
\begin{equation}
i\stackrel{\_}{\psi }\gamma ^\mu \left( \stackrel{\leftarrow }{\partial }%
_\mu -ieA_\mu \right) +m\stackrel{\_}{\psi }=0.  \label{a70}
\end{equation}

These are the equations of motions with full gauge freedom. It can be seen,
from Eq. (\ref{a63}), that $A^0$ is an arbitrary (gauge dependent) variable
since its time derivative is arbitrary. Besides that, Eq. (\ref{a62}) shows
the gauge dependence of $A^i$ and, taking the curl of its vector form, leads
to the known Maxwell equation 
\begin{equation}
\frac{\partial \stackrel{\rightarrow }{A}}{\partial t}=-\stackrel{%
\rightarrow }{E}-\stackrel{\rightarrow }{\bigtriangledown }\left( A_0-\alpha
\right) \Rightarrow \frac{\partial \stackrel{\rightarrow }{B}}{\partial t}=-%
\stackrel{\rightarrow }{\bigtriangledown }\times \stackrel{\rightarrow }{E}.
\label{a71}
\end{equation}

Writing $j^\mu =e\stackrel{\_}{\psi }\gamma ^\mu \psi $ we get , from Eq. (%
\ref{a64}), the inhomogeneous Maxwell equation 
\begin{equation}
\frac{\partial \stackrel{\rightarrow }{E}}{\partial t}=\stackrel{\rightarrow 
}{\bigtriangledown }\times \stackrel{\rightarrow }{B}-\stackrel{\rightarrow 
}{j},  \label{a72}
\end{equation}
while the other inhomogeneous equation 
\begin{equation}
\stackrel{\rightarrow }{\bigtriangledown }\cdot \stackrel{\rightarrow }{E}%
=j^0  \label{a73}
\end{equation}
follows from the secondary constraint (\ref{a46}). Expressions (\ref{a68})
and (\ref{a70}) are the known Dirac's equations for the spinor fields $\psi $
and $\stackrel{\_}{\psi }$.

\subsection{Hamilton-Jacobi formalism}

Now we apply the formalism presented in the previous sections. From the
momenta definition we have the ``{\it Hamiltonians}'' 
\begin{equation}
H_0^{\prime }=\int \left( P_0+{\cal H}_C\right) d^3x,  \label{a74}
\end{equation}
\begin{equation}
H_1^{\prime }=\int p_0d^3x,  \label{a75}
\end{equation}
\begin{equation}
H_2^{\prime }=\int \left( p_\psi -i\stackrel{\_}{\psi }\gamma ^0\right) d^3x,
\label{a76}
\end{equation}
\begin{equation}
H_3^{\prime }=\int p_{\stackrel{\_}{\psi }}d^3x,  \label{a77}
\end{equation}
which are associated, respectively, to $t=t^0$ (remember that $P_0$ is the
momentum conjugated to $t$), $A_0$, $\psi $ and $\stackrel{\_}{\psi }$. The
first two $H^{\prime }$ are even variables, while the last two are odd.
Then, using Eq. (\ref{a26}), we have 
\begin{equation}
dA^i=\frac{\delta H_0^{\prime }}{\delta p_i}dt=\left( -p^i+\partial
^iA_0\right) dt.  \label{a78}
\end{equation}

From equation (\ref{a30}) we have 
\begin{equation}
dp^i=-\frac{\delta H_0^{\prime }}{\delta A_i}dt=\left( \partial _jF^{ji}-e%
\stackrel{\_}{\psi }\gamma ^i\psi \right) dt,  \label{a79}
\end{equation}
\begin{equation}
dp^0=-\frac{\delta H_0^{\prime }}{\delta A_0}dt=\left( \partial _jp^j-e%
\stackrel{\_}{\psi }\gamma ^0\psi \right) dt,  \label{a80}
\end{equation}
\begin{equation}
dp_\psi =-\frac{\delta _rH_0^{\prime }}{\delta \psi }dt=\left( -i\partial _j%
\stackrel{\_}{\psi }\gamma ^j-e\stackrel{\_}{\psi }\gamma ^\mu A_\mu -m%
\stackrel{\_}{\psi }\right) dt,  \label{a81}
\end{equation}
\begin{equation}
dp_{\stackrel{\_}{\psi }}=-\frac{\delta _rH_0^{\prime }}{\delta \stackrel{\_%
}{\psi }}dt+-\frac{\delta _rH_2^{\prime }}{\delta \stackrel{\_}{\psi }}d\psi
=\left( -i\gamma ^j\partial _j\psi +eA_\mu \gamma ^\mu \psi +m\psi \right)
dt-i\gamma ^0d\psi .  \label{a82}
\end{equation}

The integrability conditions require $dH^{\prime }=0$, which implies for $%
H_1^{\prime }$ 
\begin{equation}
dH_1^{\prime }=d\int p_0d^3x=0\Rightarrow dp_0=\partial _jp^j-e\stackrel{\_}{%
\psi }\gamma ^0\psi =0,  \label{a83}
\end{equation}
where we made use of Eq. (\ref{a80}). This expression is equivalent to the
secondary constraint (\ref{a46}) and has to satisfy 
\begin{equation}
dH_4^{\prime }=0;\ H_4^{\prime }=\int \left( \partial _jp^j-e\stackrel{\_}{%
\psi }\gamma ^0\psi \right) d^3x,  \label{a84}
\end{equation}
which is indeed identically satisfied.

For $H_2^{\prime }$ we have 
\begin{equation}
dH_2^{\prime }=d\int \left( p_\psi -i\stackrel{\_}{\psi }\gamma ^0\right)
d^3x=0\Rightarrow dp_\psi =i\left( d\stackrel{\_}{\psi }\right) \gamma ^0,
\label{a85}
\end{equation}
which can't be written as an expression like $H^{\prime }=0$ due to the
presence of two differentials ($dp_\psi $ and $d\stackrel{\_}{\psi }$) but,
substituting in Eq. (\ref{a81}), we get 
\begin{equation}
i\left( d\stackrel{\_}{\psi }\right) \gamma ^0=\left( -i\partial _j\stackrel{%
\_}{\psi }\gamma ^j-e\stackrel{\_}{\psi }\gamma ^\mu A_\mu -m\stackrel{\_}{%
\psi }\right) dt,  \label{a86}
\end{equation}
i.e. 
\begin{equation}
-i\stackrel{.}{\stackrel{\_}{\psi }}\gamma ^0-i\partial _j\stackrel{\_}{\psi 
}\gamma ^j-e\stackrel{\_}{\psi }\gamma ^\mu A_\mu -m\stackrel{\_}{\psi }=0,
\label{a87}
\end{equation}
\begin{equation}
i\stackrel{\_}{\psi }\gamma ^\mu \left( \stackrel{\leftarrow }{\partial }%
_\mu -ieA_\mu \right) +m\stackrel{\_}{\psi }=0.  \label{a88}
\end{equation}

For $H_3^{\prime }$ we have 
\begin{equation}
dH_3^{\prime }=d\int p_{\stackrel{\_}{\psi }}d^3x=0\Rightarrow dp_{\stackrel{%
\_}{\psi }}=0,  \label{a89}
\end{equation}
that, similarly to the case above, can be used in Eq. (\ref{a82}) giving 
\begin{equation}
\left( -i\gamma ^j\partial _j\psi +eA_\mu \gamma ^\mu \psi +m\psi \right)
dt-i\gamma ^0d\psi =0,  \label{a90}
\end{equation}
\begin{equation}
i\left( \partial _\mu +ieA_\mu \right) \gamma ^\mu \psi -m\psi =0.
\label{a91}
\end{equation}

Finally, we can verify, using the above results, that $dH_0^{\prime }=0$ is
identically satisfied.

Equations (\ref{a88}) and (\ref{a91}) are identical, respectively, to
equations (\ref{a70}) and (\ref{a68}) obtained in Hamiltonian formalism.
Besides that, equations (\ref{a79}) and (\ref{a64}) are equivalent, while
Eq. (\ref{a78}) corresponds to Eq. (\ref{a62}) except for an arbitrary gauge
factor.

\section{Conclusions}

In this work we presented a formal generalization of the Hamilton-Jacobi
formalism for singular systems with Berezinian variables, obtaining the
equations of motion as total differential equations (\ref{a26}) and (\ref
{a30}). In these equations, the coordinates $q^\alpha =t^\alpha $ ($\alpha
=1,...,R$), whose momenta are constrained, play the role of evolution
parameters of the system, together with the time parameter $t=t^0$. So, the
system's evolution is described by contact transformations generated by the
``Hamiltonians'' $H_\alpha ^{\prime }$ and parametrized by $t^\alpha $ (with 
$\alpha =0,1,...,R$), were $H_0^{\prime }$ is related to the canonical
Hamiltonian by Eq. (\ref{a22}) and the other $H_\alpha ^{\prime }$ ($\alpha
=1,...,R$) are the constraints given by Eq. (\ref{a23}). This evolution is
considered as being always restricted to the constraints surface in phase
space, there is no complete phase space treatment that is latter reduced to
the constraints surface, as in Dirac's formalism with the use of weak
equalities.

We should observe that, in the case of systems composed exclusively by even
variables, all parities are equal to zero and equations (\ref{a26}), (\ref
{a30}), (\ref{a31}) reduce to the results obtained in ref.\cite{Guler 1}.
Furthermore, if the system is nonsingular, we have $H_\alpha ^{\prime
}\equiv 0$ except for $\alpha =0$, so the total differential equations (\ref
{a26}) and (\ref{a30}) will be reduced to the expressions given by Eq. (\ref
{a4}).

The integrability conditions (which relation to the consistency conditions
in Dirac's formalism is discussed in the appendix) were shown to be
equivalent to the necessity of the vanishing of the variation of each $%
H_\alpha ^{\prime }$ ($\alpha =0,...,R$), i.e. $dH_\alpha ^{\prime }=0$.

The example presented was chosen for its completeness: it is a singular
system with even and odd variables and its Hamiltonian treatment contains
all kinds of constraints (primary and secondary, first and second class
ones). This example is very illustrative, since it allows a comparison
between all features of Dirac's and Hamilton-Jacobi formalisms. For example,
the fact that the integrability conditions $dH_2^{\prime }=0$ and $%
dH_3^{\prime }=0$ give expressions involving some differentials $dt^\alpha $
is related to the fact that the corresponding Hamiltonian constraints $\phi
_2$ and $\phi _3$ are second class constraints and determine some of the
arbitrary parameters in the primary Hamiltonian (\ref{a39}). Similarly, the
fact that the condition $dH_1^{\prime }=0$ generated an expression like $%
H_4^{\prime }=0$ is related to the fact that the corresponding Hamiltonian
constraint $\phi _1$ is a first class one (see appendix).

Finally, we must call attention to the presence of arbitrary variables in
some of the Hamiltonian equations of motion due to the fact that we have
gauge dependent variables and we have not made any gauge fixing. This does
not occur in Hamilton-Jacobi formalism since it provides a gauge-independent
description of the systems evolution due to the fact that the
Hamilton-Jacobi function $S$ contains all the solutions that are related by
gauge transformations.

\section{Appendix: Equivalence among consistency \newline and integrability conditions
}

In this appendix we will show the equivalence among the integrability
conditions of the formalism showed above and the consistency conditions in
Dirac's Hamiltonian formalism, in a similar way to what was made for usual
variables \cite{Rabei}. In the notation used in this paper the Dirac's
primary constraints are written, from Eq. (\ref{a23}), as 
\begin{equation}
H_\alpha ^{\prime }\equiv P_\alpha +H_\alpha \left( q^i;p_a\right) \approx 0,
\label{ap1}
\end{equation}
where $\alpha =1,...,R$; $i=1,...,N$. The canonical Hamiltonian is given by $%
H_0$ in Eq. (\ref{a20}), so the primary Hamiltonian $H_P$ is 
\begin{equation}
H_P\equiv H_0+H_\alpha ^{\prime }v^\alpha ,  \label{ap2}
\end{equation}
where the $v^\alpha $ are unknown coeficients related to the undetermined
velocities $\stackrel{.}{q}^\alpha $ \cite{Sundermeyer}. The ordering of the 
$v^\alpha $ with respect to the $H_\alpha ^{\prime }$ is a matter of choice,
since it will simply produce a change of sign, but the natural procedure,
that identifies $v^\alpha $ and $\stackrel{.}{q}^\alpha $, suggests the
ordering above as a consequence of the ordering adopted in the Hamiltonian (%
\ref{a3}). This ordering is also the most natural choice to our purpose but
is, of course, irrelevant for systems containing only usual variables. The
consistency conditions, which demand that the constraints preserved by time
evolition, are written as 
\begin{equation}
\stackrel{.}{H}_\mu ^{\prime }\approx \left\{ H_\mu ^{\prime },H_P\right\}
_B=\left\{ H_\mu ^{\prime },H_0\right\} _B+\left\{ H_\mu ^{\prime },H_\alpha
^{\prime }\right\} _B\stackrel{.}{q}^\alpha =0,  \label{ap3}
\end{equation}
where $\alpha ,\mu =1,...,R$ and the Berezin brackets here are that given in
Eq. (\ref{a5}) defined in the usal $2N$ dimensional phase space and we have
made the explicity identification $\stackrel{.}{q}^\alpha \equiv v^\alpha $.

Multipling the above equation by $dt$ we get 
\begin{equation}
dH_\mu ^{\prime }\approx \left\{ H_\mu ^{\prime },H_0\right\} _Bdt+\left\{
H_\mu ^{\prime },H_\alpha ^{\prime }\right\} _Bdt^\alpha =0,  \label{ap4}
\end{equation}
where, as before, $q^\alpha =t^\alpha $ but we are still making $\alpha
=1,...,R$. At this point we can already see that, when Dirac's consistency
conditions are satisfied we have $dH_\mu ^{\prime }=0$ satisfied. We must
see now that we have $dH_0^{\prime }=0$ when Dirac's consistency conditions
are satisfied. The Hamiltonian equation of motion for $H_0^{\prime }$ is 
\begin{equation}
\stackrel{.}{H}_0^{\prime }\approx \left\{ H_0^{\prime },H_P\right\}
_B=\left\{ H_0^{\prime },H_0^{\prime }\right\} _B+\left\{ H_0^{\prime
},H_\alpha ^{\prime }\right\} _B\stackrel{.}{q}^\alpha ,  \label{ap5}
\end{equation}
which, multiplied by $dt$ becomes 
\begin{equation}
dH_0^{\prime }\approx \left\{ H_0^{\prime },H_0\right\} _Bdt+\left\{
H_0^{\prime },H_\alpha ^{\prime }\right\} _Bdt^\alpha .  \label{ap6}
\end{equation}

Remembering that the ``momentum'' $P_0$ in $H_0^{\prime }$ is independent of
the canonical variables $q^i$ and $p_i$, we have 
\begin{equation}
dH_0^{\prime }\approx \left\{ H_0,H_0\right\} _Bdt+\left\{ H_0,H_\alpha
^{\prime }\right\} _Bdt^\alpha =\left\{ H_0,H_\alpha ^{\prime }\right\}
_Bdt^\alpha .  \label{ap7}
\end{equation}

But, if Dirac's consistency conditions are satisfied, we must have only
primary first class constraints, otherwise we would have conditions imposed
on the unknown velocities $\stackrel{.}{q}^\alpha $. So, the preservation of
constraints in time will reduce to 
\begin{equation}
\left\{ H_\mu ^{\prime },H_0\right\} _B\approx 0,  \label{ap8}
\end{equation}
and the right side of Eq. (\ref{ap7}) will be zero. This is simply a
consequence of the fact that, once all Dirac's conditions are satisfied, the
Hamiltonian is preserved. So the condition $dH_0^{\prime }=0$ is satisfied
when Dirac's consistency conditions are satisfied.

This shows that the integrability conditions in Hamilton-Jacobi formulation
will be satisfied when Dirac's consistency are satisfied. Similarly, we can
consider that we have the integrability conditions satisfied so that $%
dH_0^{\prime }=dH_\mu ^{\prime }=0$ and then Eq. (\ref{ap4}), which is
equivalent to Eq. (\ref{ap3}), implies that Dirac's conditions are
satisfied. So, both conditions are equivalent.

Now, we will consider that these conditions are not initialy satisfied. When
we have only first class constraints in Hamiltonian formalism we will simply
get a new constraint from some of the conditions (\ref{ap3}). From Eq. (\ref
{ap4}) we see that this will imply an expression like $dH_\mu ^{\prime
}=H^{\prime }dt$ ($H^{\prime }\approx 0$ is the secondary Hamiltonian
constraint) which means that there will be a new $H^{\prime }$ in
Hamilton-Jacobi formalism that have to satisfy $dH^{\prime }=0$.

If we have some second class Hamiltonian constraints the consistency
conditions (\ref{ap3}) will imply a condition over some of the velocities $%
\stackrel{.}{q}^\alpha $. From Eq. (\ref{ap4}) we see that, in
Hamilton-Jacobi approach, there will be conditions imposed on some
differentials $dt^\alpha $.

Such correspondence among the formalisms can be clearly seen in the example
presented in this paper.

Besides that, Eq. (\ref{ap4}) and Eq. (\ref{ap6}) can be written as 
\begin{equation}
dH_\mu ^{\prime }\approx \left\{ H_\mu ^{\prime },H_\alpha ^{\prime
}\right\} _Bdt^\alpha ,  \label{ap9}
\end{equation}
were now $\alpha ,\mu =0,1,...,R$ and the Berezin bracket is again defined
in the $2N+2$ dimensional phase space containing $t^0$ and $P_0$. This
equation is obviously identical to Eq. (\ref{i14}), that leads to the
integrability condition $dH_\mu ^{\prime }=0$, and its right hand side was
showed to correspond to Dirac's consistency conditions. Consequently, this
expression shows directly the relation among consistency and integrability
conditions.

It's important to notice that here we are not considering any explicity
dependence on time, neither of the constraints nor of the canonical
Hamiltonian, because it is an usual procedure in Hamiltonian approach. But
the equations of Hamilton-Jacobi formalism were obtained without considering
this condition and, consequently, remain valid if we consider systems with
Lagrangians that are explicitily time dependent.

But Hamiltonian approach is also applicable to such systems (see reference 
\cite{Gitman}, page 229) and in this case we can follow a procedure similar
to that one showed here and demonstrate the correspondence among Dirac's
consistency conditions and integrability conditions.

Finally, some words about the simpletic structure. Using Eq. (\ref{i8}), we
can writte Eq. (\ref{a26}) and Eq. (\ref{a30}) in terms of left derivatives
as 
\begin{equation}
dq^i=\frac{\partial _lH_\alpha ^{\prime }}{\partial p_i}dt^\alpha ,
\label{ap10}
\end{equation}
\begin{equation}
dp_i=-\left( -1\right) ^{P_{\left( i\right) }}\frac{\partial _lH_\alpha
^{\prime }}{\partial q^i}dt^\alpha ,  \label{ap11}
\end{equation}
where, as before, $i=0,1,...,N$ and $\alpha =0,1,...,R$. These expressions
can be compactly written as 
\begin{equation}
d\eta ^I=E^{IJ}\frac{\partial _lH_\alpha ^{\prime }}{\partial \eta ^J}%
dt^\alpha ,  \label{ap12}
\end{equation}
were we used the notation 
\begin{equation}
\eta ^{1i}=q^i,\ \eta ^{2i}=p_i\Rightarrow \eta ^I=\left( q^i,p_i\right) ,\
I=\left( \zeta =1,2;\ i=0,1,...,N\right) ,  \label{ap13}
\end{equation}
\begin{equation}
E^{IJ}=\delta ^i{}_j\left( \delta ^\zeta {}_1\delta ^2{}_\sigma -\left(
-1\right) ^{P_{\left( i\right) }}\delta ^1{}_\sigma \delta ^\zeta
{}_2\right) ,\ I=\left( \zeta ;\ i\right) ,\ J=\left( \sigma ;\ j\right) ;
\label{ap14}
\end{equation}
that was introduced in page 76 of reference \cite{Gitman}. The Berezin
brackets defined in Eq. (\ref{a5}) can be written as 
\begin{equation}
\left\{ F,G\right\} _B=\frac{\partial _rF}{\partial \eta ^I}E^{IJ}\frac{%
\partial _lF}{\partial \eta ^J}.  \label{ap15}
\end{equation}

This simpletic notation allows us to obtain the expression for the total
differential for any function $F\left( t^\beta ,q^a,p^a\right) $ in a more
direct way. Using it in Eq. (\ref{i12a}) we get 
\begin{equation}
dF=\frac{\partial _rF}{\partial \eta ^I}d\eta ^I,  \label{ap16}
\end{equation}
where the use of Eq. (\ref{ap12}) gives 
\begin{equation}
dF=\frac{\partial _rF}{\partial \eta ^I}E^{IJ}\frac{\partial _lH_\alpha
^{\prime }}{\partial \eta ^J}dt^\alpha =\left\{ F,H_\alpha ^{\prime
}\right\} _Bdt^\alpha ,  \label{ap17}
\end{equation}
in agreement with Eq. (\ref{i13}).

\section{Acknowledgments}

B. M. P. is partially supported by CNPq and R. G. T. is supported by CAPES.


\newpage 

\begin{thebibliography}{99}
\bibitem{Guler 1}  Y. G\"{u}ler, {\it Il Nuovo Cimento} B {\bf 107} (1992),
1389.

\bibitem{Guler 2}  Y. G\"{u}ler, {\it Il Nuovo Cimento} B {\bf 107} (1992),
1143.

\bibitem{Gitman}  D. M. Gitman and I. V. Tyutin, ``Quantization of Fields
with Constraints,'' Springer-Verlag, 1990.

\bibitem{Sundermeyer}  K. Sundermeyer, ``Lecture Notes in Physics 169 -
Constrained Dynamics,'' Springer-Verlag, 1982.

\bibitem{Regge}  A. Hanson, T. Regge and C. Teitelboim, ``Constrained
Hamiltonian Systems,'' Accademia Nazionale dei Lincei, Roma, 1976.

\bibitem{Eu 1}  B. M. Pimentel and R. G. Teixeira, {\it Il Nuovo Cimento} B 
{\bf 111} (1996), 841.

\bibitem{Eu 2}  B. M. Pimentel and R. G. Teixeira, {\it Preprint
hep-th/9704088}, to appear in {\it Il Nuovo Cimento} B.

\bibitem{Guler 4}  Y. G\"{u}ler, {\it Il Nuovo Cimento} B {\bf 109} (1994),
341.

\bibitem{Guler 6}  Y. G\"{u}ler, {\it Il Nuovo Cimento} B {\bf 111} (1996),
513.

\bibitem{Berezin}  F. A. Berezin, ``Introduction to Superanalysis,'' D.
Reidel Publishing Company, Dordrecht, Holland, 1987.

\bibitem{Caratheodory}  C. Carath\'{e}odory, ``Calculus of Variations and
Partial Differential Equations of the First Order,'' Part II, p. 205,
Holden-Day, 1967.

\bibitem{Mukunda}  E. C. G. Sudarshan and N. Mukunda, ``Classical Dynamics:
A Modern Perspective,'' John Wiley \& Sons Inc., New York, 1974.

\bibitem{Guler 3}  E. M. Rabei and Y. G\"{u}ler, {\it Phys. Rev.} A {\bf 46}
(1992), 3513.

\bibitem{Guler 5}  Y. G\"{u}ler, {\it Il Nuovo Cimento} B {\bf 110} (1995),
307.

\bibitem{Whittaker}  E. T. Whittaker, ``A Treatise on the Analytical
Dynamics of Particles and Rigid Bodies, '' 4th ed., p. 52, Dover, 1944.

\bibitem{Rabei}  E. M. Rabei, {\it Hadronic Journal} {\bf 19} (1996), 597.
\end{thebibliography}
\end{document}